\documentclass[twocolumn,showpacs,preprintnumbers,amsmath,amssymb,prl]{revtex4}

\usepackage{graphicx}
\usepackage{dcolumn}
\usepackage{bm}
\usepackage{setspace}

\begin{document}

\title{Manipulating the magnetic structure by electric fields in multiferroic ErMn$_{2}$O$_{5}$}

\author{Y. Bodenthin$^1$}
\author{U. Staub$^1$}
\author{M. Garc\'{i}a-Fern\'{a}ndez$^1$}
\author{M. Janoschek$^2$}
\author{J. Schlappa$^1$}
\author{E. I. Golovenchits$^3$}
\author{V. A. Sanina$^3$}
\author{S. G. Lushnikov$^3$}

\address{$^1$Swiss Light Source, Paul Scherrer Institut, CH-5232
Villigen PSI, Switzerland
  \\$^2$Laboratory for Neutron
Scattering, ETH Zurich and Paul Scherrer Institut, CH-5232 PSI,
Villigen
\\$^3$Ioffe Physical Technical Institute, 26 Politekhnicheskaya,
194021 St. Petersburg, Russia }

\date{\today}

\begin{abstract}
Based on measurements of soft x-ray magnetic diffraction under
in-situ applied electric field, we report on significantly
manipulation and exciting of commensurate magnetic order in
multiferroic ErMn$_2$O$_5$. The induced magnetic scattering
intensity arises at the commensurate magnetic Bragg position
whereas the initial magnetic signal almost persists. We
demonstrate the possibility to imprint a magnetic response
function in ErMn$_2$O$_5$ by applying electric field.
\end{abstract}

 \maketitle

\begin{figure}
\includegraphics[width=8.4cm,height=4.2cm]{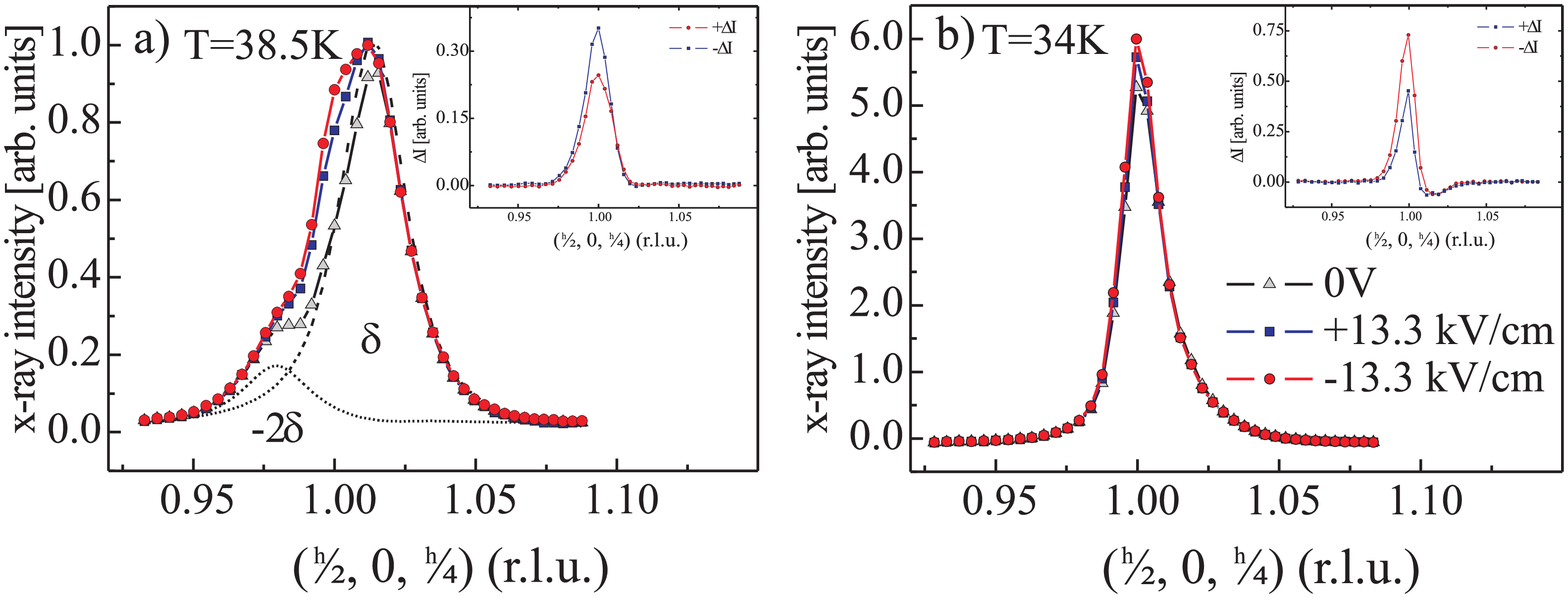}
\caption{\label{fig1} $\theta/ 2 \theta$ scans of the
($\frac{1}{2}$,0,$\frac{1}{4}$)-reflection of ErMn$_2$O$_5$ taken
at the Mn $L_{3}$-edge at $T=34K$ below the 2D to 1D-ICM phase
transition (a) and above at $T=38.5K$ (b) with $\pi$-polarized
incident light. $\delta$ corresponds
to the magnetic and $-2\delta$ to the induced aspheric charge
reflection (see text). The inset show the intensity difference $\Delta I^{\pm}= I(\pm E)-I(0V)$.}
\end{figure}

\begin{figure}
\includegraphics[width=6cm,height=8.1cm]{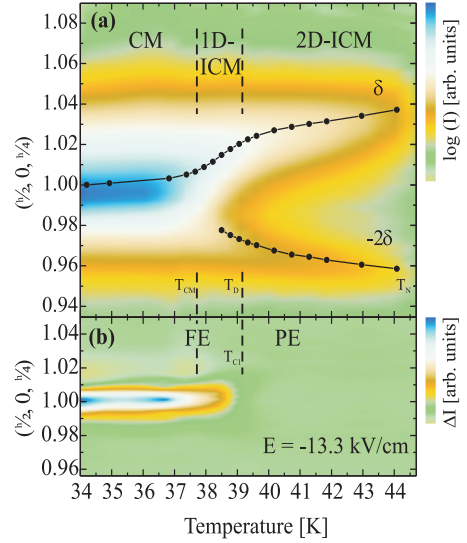}
\caption{\label{fig2}(a): Temperature dependence of the
magnetic $(\frac{1}{2}, 0 , \frac{1}{4})$-reflection and the
incommensurate $\delta$ and $-2\delta$ reflections of
ErMn$_2$O$_5$. $T_D = T_{C1} =39.1K$ is defined by the change in
slope of $\delta(T)$ and $-2\delta(T)$. At $T_{CM}=37.7K$,
ErMn$_2$O$_5$ enters
the commensurate magnetic phase (CM). (b): Intensity difference $\Delta I(T)$ as function
of temperature. The appearance of $\Delta I$ at $T_{D}$ is
explicitly shown in Figure 3c.}
\end{figure}

\begin{figure}
\includegraphics[width=8.6cm,height=6.3cm]{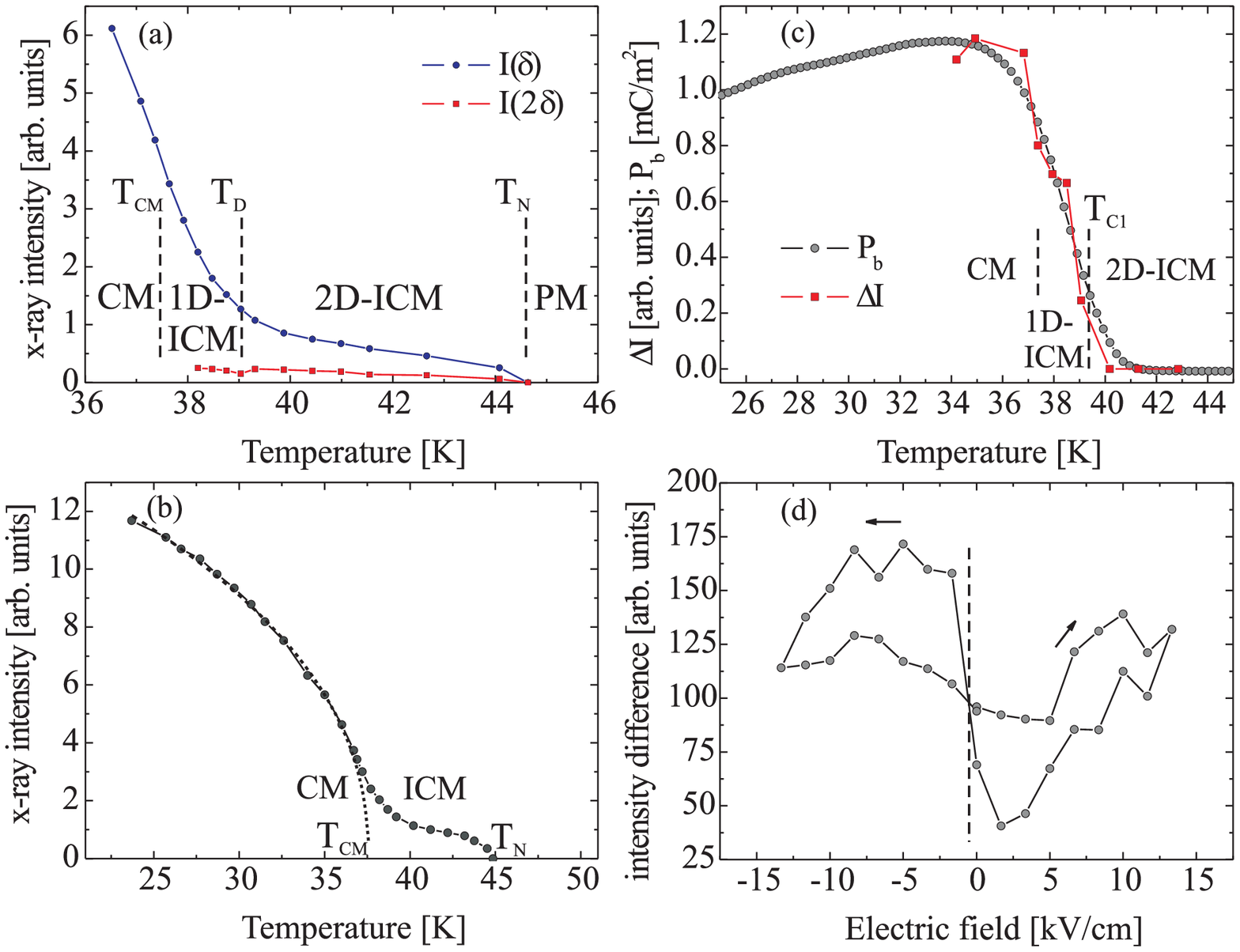}
\caption{\label{fig3}(a): X-ray intensity $I$ as function of
temperature for the $\delta$ and $-2\delta$ satellites. (b): The
temperature dependence of $I$ (solid line and dots) for $\delta$
follows a power law at low temperatures (dotted line). (c):
$\Delta I(T)$ follows the polarization $P_b$ for $H=0T$
(reproduced from Ref. \cite{higashiyama}). (d): X-ray intensity
$I$ as function of the applied electric field at $T=38.5K$,
showing its hysteretic behavior after FC(+). The dotted line shows
the bias field.}
\end{figure}

The coexistence of magnetism and ferroelectricity in solid
materials is unusual, and still more unusual is the coupling
between these phenomena. Materials with such couplings are often
called multiferroics, and the coupling is based on the
magneto-electric effect. The interest in these materials lies in
the fact that a manipulation of electric (magnetic) properties by
magnetic (electric) fields would be of great use in spintronic
devices \cite{fiebig,tokura,cheong}. In recent years, it has been
shown that non-collinear, long-range antiferromagnetic structures
can break spatial inversion symmetry and
drive a ferroelectric modification \cite{kimura_nature,Hur}. Ferroelectricity may arise regardless of the
commensurability between the lattice and magnetic order
\cite{chaponandradelli}. Electric polarization produced by the
spin current (vector spin chirality) $\mathbf{S_i \times S_{j}}$
may even arise in the presence of a center of inversion symmetry
between spins, i.e., without Dzyaloshinskii-Moriya interaction,
with a noncollinear configuration such as the transverse-spiral
and the spontaneous polarization normal to the spiral propagation
vector
\cite{DM,KNB}. \\
Early work on manipulating magnetization by an
electric field was reported by \textit{Ascher et al.}, who
observed that a reversal of the electric polarization $\mathbf{P}$
of Ni$_3$B$_7$O$_{13}$I from the $[001]$ crystallographic axis to
$[00 \overline{1}]$ by an electric field leads to a rotation of
the weak ferromagnetic moment $\mathbf{m}$ from $[110]$
to $[1 \overline{1}0]$ \cite{ascher}. Much later, it was
demonstrated that an electric field controls the magnetic order of
the Ho$^{3+}$-ions in hexagonal HoMnO$_3$ and that
antiferromagnetic domain patterns correlate with ferroelectric
domains in thin BiFeO$_3$ films \cite{lottomoser,zhao}. Recently,
an elegant way to control the spin helicity by an electric field
was discovered in the magnetic spiral compound TbMnO$_3$
\cite{yamasaki1}: the reversal of spin helicity was achieved by
the electric field cooling through $T_C$. Nevertheless, the
coercive electric field was too large $(>20 kV/cm)$ to reverse the
ferroelectric polarization below the ferroelectric Curie
temperature. Despite these few studies, little is known about how magnetic structures are influenced by electric fields.\\
In the \textit{R}Mn$_2$O$_5$ series with
\textit{R}=Ho, Tb, Dy, Y and Er the magnetoelectric couplings are
gigantic, and the magnetic phases involved are complicated and
commonly incommensurate with lattice \cite{chapon,chapon2,cruz,
Hur}. The spontaneous electric polarization $\mathbf{P}$ results from
acentric spin-density waves \cite{chapon2} whereas little is known
about the underlying mechanism of multiferroicity because of their structural and magnetic complexity.
Furthermore, the brake of inversion symmetry in the ferroelectric
magnetic phases implies a coupling of magnetism to odd orders
of $\mathbf{P}$ \cite{kenzelmann_symmetry,arima_symmetry}.\\
ErMn$_2$O$_5$ shows spontaneous electric
polarization along the $b$-axis at $T_{C1}=39.1K$, i.e. below the
N\'{e}el temperature of $T_{N1}=44K$ \cite{kobayashi,fukunaga}.
The system enters a commensurate magnetic (CM) phase at
$T_{CM}=37.7K$ with a magnetic ordering vector of
$\mathbf{q}=(\frac{1}{2},0,\frac{1}{4}$) upon cooling from a
two-dimensionally-modulated incommensurate magnetic structure
(2D-ICM) with $\mathbf{q}=(q_x,0,q_z$). Thereby the system passes
through a one-dimensionally-modulated incommensurate magnetic
phase (1D-ICM) with $\mathbf{q}=(q_x,0,\frac{1}{4}$). The
transition temperature $T_{D}$ between the two incommensurate
magnetic phases is correlated to the Curie temperature of the
spontaneous electric polarization at $T_{C1}=T_{D}=39.1K$. In the
present paper, we demonstrate that ferroelectricity in
ErMn$_2$O$_5$ strongly couples with the commensurate magnetic
structure and that the magnetic order can be deliberately
modulated, excited and switched by applying a static electric
field.\\
ErMn$_2$O$_5$ single crystals were grown by spontaneous
crystallization \cite{einkristall}. Resonant soft x-ray magnetic
diffraction was performed using the RESOXS station at the SIM
beamline of the Swiss Light Source at the Paul Scherrer Institut,
Switzerland \cite{Staub_Resoxs}. The incoming linearly polarized
light was oriented either parallel ($\pi$) or perpendicular
($\sigma$) to the scattering plane. The sample was mounted on a
sapphire plate on a continuous-helium-flow cryostat, providing a
temperature range of $29K \leq T \leq 300 K$. The $b$ axis was aligned parallel to the scattering plane,
and the electric field was applied
perpendicular to both the $b$ axis and the magnetic wave vector ($\frac{1}{2}$,0,$\frac{1}{4}$).\\
In Figure \ref{fig1} we present resonant magnetic soft x-ray
diffraction of ErMn$_2$O$_5$ taken with $\pi$-polarized light by
$\theta/2\theta$ scans at the Mn $L_{3}$-edge ($E=643.75 eV$; probing the Mn moments only) at
two different temperatures, below (Figure 1a) and above the ICM-CM
phase transition, respectively.
Magnetic Bragg scattering appears below $T_{CM}$ at
$\mathbf{q}=(\frac{1}{2}$,0,$\frac{1}{4}$), whereas above $T_{CM}$
two satellite reflections $\delta$ and $-2\delta$ appear. $\delta$
denotes the deviation in $\mathbf{q}$ from the commensurate
values. The resonant magnetic scattering amplitude relevant for the antiferromagnetic reflection in the electric
dipole ($E1$) approximation can be written as
\begin{equation}
f^{res}_{E1}\propto i(\mathbf{e}_i \times
\mathbf{e}_{f}^{*})\cdot \mathbf{m} F^{(1)}
+(\mathbf{e}_i \cdot \mathbf{m})(\mathbf{e}_{f}^{*} \cdot
\mathbf{m})F^{(2)},
\end{equation}
were $\mathbf{m}$ denotes the local moment direction \cite{hannon,lovesey}. The first
term depends linearly on $\mathbf{m}$ and gives first-harmonic
satellites ($\delta$), whereas the second term corresponds to
orbital scattering leading to the second-harmonic satellite $-2
\delta$ (quadratic in $\mathbf{m}$), and describes the induced
charge anisotropy of Mn which is supported by different polarization and
energy dependencies of these reflections. \\
Applying an electric field of $E = \pm 13.3kV/cm$
perpendicular to the directions of the ferroelectric polarization
and the magnetic wave vector $\mathbf{q}$ leads to a pronounced
increase of the scattered magnetic intensity at
$\mathbf{q}=(\frac{1}{2},0,\frac{1}{4})$ in the commensurable as
well in the 1D-ICM phase as shown in Figure 1a and b. We observe a distinct
different magnitude for positive and negative electric field
direction ($10\%$) which is in relation to the recent findings of the tendency of electric
polarization $\mathbf{P}$ to be spontaneously oriented in a preferred
direction \cite{fukunaga}. Therefore, we conclude this to be a general feature in
ErMn$_2$O$_5$ and not caused by extrinsic and sample dependent
effects. The intensity difference $\Delta I^{\pm}= I(\pm E) - I(0)$ for zero field cooling
(ZFC) and for different field cooled (FC) scenarios confirm this
findings (data not shown). The insets in Figure 1 show the intensity difference $\Delta I^{\pm}$ for ZFC. The small dip in
the difference intensities on the right side is likely
due to a reduction of the intensity from the incommensurate $\delta$
magnetic peak, indicating that the observed difference in the
intensity is in part due to a change from 1D-ICM to the CM phase.
Together with the asymmetric peak shape of the commensurate
reflection, this is an indication for a coexisting of phases as
observed for YMn$_{2}$O$_{5}$ and TbMn$_{2}$O$_{5}$
\cite{kobayashi2,okamoto}. These
findings show the direct evidence of manipulation and
excitation of the magnetic
structure with an in-situ applied electric field.\\
To obtain further insight, a detailed temperature dependence of
the reflection with and without an electric field was collected,
providing an in-situ measurement of the coupled magnetic and
ferroelectric transitions in ErMn$_2$O$_5$. The upper part of
Figure \ref{fig2} presents $\theta /2 \theta$-scans across the
magnetic reflection without any applied electric field in the
temperature interval $34K \leq T \leq 44.7K$.
The N\'{e}el-temperature at $T_N=44K$ is observed with the onset
of magnetic scattering. The fact that $T_N$ is identical for both
the $\delta$ and $-2 \delta$ satellites establishes that they
originate from magnetic ordering. Obviously, the magnetic spiral
(represented by $\delta$) drives the aspheric charge density wave
($-2 \delta$). Fits were used to establish the peak positions of
$\delta$ and $-2\delta$ given by the black dots in Figure
\ref{fig2}a. By further lowering the temperature to $T_{D}=39.1K$,
the positions of $\delta$ and $-2\delta$ slightly changes. With
decreasing temperature, the magnetic structure becomes
one-dimensionally-modulated incommensurate (1D-ICM) at $T_D
=39.1K$ \cite{kobayashi,fukunaga}. The lock-in of $q_{z}$ to the
incommensurate value leads to a significant change in the slope of
$\delta(T)$ and $-2\delta(T)$, solely $q_{x}$ changes further on.
The phase transition is marked by the first vertical line in
Figure \ref{fig2}. With a further decrease in temperature, ErMn$_2$O$_5$
enters the commensurate magnetic phase (CM) at $T_{CM}=37.7K$.
At this point $\delta(T)$ is indistinguishable from $-2\delta(T)$ and both reflections merge into the commensurate $(\frac{1}{2},0,\frac{1}{4})$-reflection.\\
Simultaneously, we measure the influence of the applied electric
field on the magnetic scattering at each temperature. The lower
part of Figure \ref{fig2} presents the intensity difference
$\Delta I(T)$ between a scan with and without an applied electric
field of $E = - 13.3 kV/cm$. Obviously, the onset of
$\Delta I(T)$ is associated with the 2D-ICM to 1D-ICM transition.
Since the $E$ field is applied in-situ, these results represent
direct proof of a coupling between ferroelectric and magnetic
order, since no change in the magnetic signal is observed in the
paraelectric phase. Moreover, the intensity difference peak
appears at the commensurate peak position and $\Delta I$ is stable
in $\mathbf{q}$.\\
Below $T_{C1}$, Mn spins are excited into the non-collinear
commensurate magnetic structure, with the propagation vector
$\mathbf{q}=(\frac{1}{2}, 0, \frac{1}{4}$) by the application of
an electric field. These findings demonstrate the establishment of
commensurality in the 1D-ICM phase by the presence of an electric
field and hence an influence of the magnetic moments by $E$. More information on the magnetic transitions is available from
the integrated intensities $I(T)$ for $\delta$ and $-2\delta$,
which are plotted as a function of temperature in Figure
\ref{fig3}a.
The onset of $I(T)$ coincides with the N\'{e}el temperature
($T_{N}$) for both satellite reflections. The change in
$\mathbf{q}$ for $\delta (T)$ and $-2 \delta (T)$, as observed in
Figure \ref{fig2}a, is accompanied by a significant change in
intensity when passing through the 2D-ICM to 1D-ICM (PE to FE)
transition at $T_{D}(T_{C1})$. Finally, the appearance of the
commensurate magnetic phase at $T_{CM}$ is signalled by the
merging of $\delta (T)$ and $-2 \delta (T)$ to the commensurate
peak, though the results of Figure 1 indicate that a weak ICM contribution still coexists. In Figure \ref{fig3}b, the fit of $I(T)$ at lower
temperatures to a power law is given. Extending the fit, the curve
crosses zero at approximately $T=37.6K$, indicating the onset of
the commensurate magnetic structure. Note that in case of
coexisting phases, as suggested in the discussion of Figure 1, the
fit does not reflect a critical exponent. Figure \ref{fig3}c shows
$\Delta I(T)$ as function of temperature normalized to the initial
intensity. $\Delta I(T)$ follows closely the electric
polarization $P_b$ measured by \textit{Higashima et al.} with the
onset of $\Delta I(T)$ at approximately $T_{C1}$. $P_b(T)$ is reproduced from
reference \cite{higashiyama}, with an adjustment to temperatures
measured by \emph{Kobayashi et al.} and \emph{Fukunaga et al.}
\cite{kobayashi,fukunaga}. The dependence of the intensity $\Delta
I(E)$ as function of the applied electric field is shown in Figure
\ref{fig3}d. The data were taken after FC(+) scenario which leads
to positive $\Delta I$ values at $0V$ and a negative field bias.
As reference we use the $I(0V)$ measured with ZFC on exact
the same temperature. This allows to obtain a hysteresis by
sweeping the electric field $E$. We observe an increasing of
$\Delta I$ when increasing $E$ up to $\sim 9 kV/cm$, followed by a
decreasing of the intensity difference by further increasing the
applied electric field up to the maximum value of $E=13.3 kV/cm$.
Similar behavior could also be observed by applying the electric
field in the opposite direction as displayed in the left part of
the hysteresis. A second remarkable finding is the difference in
the slope of the hysteresis for increasing and decreasing $E$ when
passing the $0V$ position after turning the field. It shows that
the system is in a different magnetic state as the response to the
$E$ field depends significantly on history, despite the fact that the
small negative bias already proves the imprint of a magnetic
response. The decrease of $\Delta I$ for $E> \pm 9 kV/cm$ is likely
the origin of the appearance of the hysteresis and causes the
switching of the magnetic states. One would therefore expect, that
smaller $E$ fields, though still leading to a change of intensity,
show no hysteresis. Earlier studies seems to support this
assumption as they reveal that ErMn$_{2}$O$_{5}$ shows an unusual
magnetic field dependence of magnetoelectric polarization leading
to complicated \emph{P-H} hysteresis, whereas only high order terms up
to the fourth order could describe the results satisfactory
\cite{ferroelectrics1,ferroelectrics2}. Our measurements of the
hysteretic behavior of the magnetic Bragg intensity shows a clear
memory effect of $\Delta I (E)$. Moreover, the slope of the
hysteresis depends significantly on history. Nevertheless, this
measurement is a clear indication that the magnetism in
ErMn$_2$O$_5$ can be switched between two different states by the
electric field. Additionally, a difference in $\Delta I$ of
approximately $15.3\%$ is observed between the FC(+) and FC(-)
field cooled scenarios.
\\
Finally, we probe the polarization dependency of the intensity difference $\Delta I$. For $0V$, we find the intensity ratio
between $\sigma$ and $\pi$ polarized light to be $\pi / \sigma
|_{0V}=2.163$. Applying an electric field leads to ratios of $\pi
/ \sigma |_{- \Delta E}=2.08$ and $\pi / \sigma |_{+ \Delta
E}=2.91$. Since $\sigma\sigma$-scattering is absent for magnetic
scattering and assuming the contribution of the orbital scattering
to be small, the change in the intensity ratios is an indication
that the direction of the magnetic moments is changed by applying
an electric field rather than a simple enhancement of the
moments. Based on the orthogonal character of the DM interaction,
one would assume that an electric field applied perpendicular to both,
the direction of the ferroelectric polarization and $\mathbf{q}$,
would induce a magnetic moment along the crystallographic b axis.
Since the manganese magnetic moments have components along all
three crystallographic directions \cite{chapon2} and considering
that the magnetic structure is ambiguous, the influence on the
structure factor cannot be determined quantitatively. However,
this significant change observed in the polarization ratio
indicates that the electric field rotates the magnetic moments
leading to an electric field
dependent commensurate magnetic structure.\\
In summary, resonant magnetic soft x-ray diffraction experiments
were performed on multiferroic ErMn$_2$O$_5$. Applying a static
electric field leads to a significant increase of the magnetic
scattering intensity. The difference in scattered intensity
clearly demonstrates the generation of magnetic scattering
intensity at the commensurate
$(\frac{1}{2},0,\frac{1}{4})$-position which is stable in
$\mathbf{q}$. The appearance of intensity difference $\Delta I
(T)$ as function of temperature reveals the coincidence of the
2D-ICM to 1D-ICM magnetic transition and the para- to
ferroelectric transition. In the ferroelectric phase, an applied
electric field pushes the system into the commensurate magnetic
phase by changing the direction of the magnetic moments.
Hysteresis loops as well as ZFC and FC experiments reveal the
possibility to imprint a magnetic response function by an electric
field.
\begin{acknowledgments}
We have benefited from valuable discussions with S. Lovesey, S.
Gvasaliya and B. Roessli and from the experimental support of the
X11MA beamline staff. The work was partially supported by the by
RFBR grants 05-02-17822 and 05-02-16328 and by Presidium of
Russian Academy of Sciences grant P3. The financial support of the
Swiss National Science Foundation is grateful acknowledged.
\end{acknowledgments}

\bibliography{ErMn2O5}

\end{document}